\documentclass[a4paper,UKenglish]{lipics}
 
\usepackage{microtype}

\title{Euler diagrams as an introduction to set-theoretical models
}

\author{Ryo Takemura}
\affil{Nihon University\\
5-2-1 Kinuta, Setagaya-ku, Tokyo 157-8570, Japan\\
  \texttt{takemura.ryo@nihon-u.ac.jp}}
\authorrunning{R. Takemura} 

\Copyright{Ryo Takemura}


\serieslogo{logo_ttl}
\volumeinfo
  {M. Antonia {Huertas}, Jo\~ao {Marcos}, Mar\'ia {Manzano}, Sophie {Pinchinat}, \\
  Fran\c{c}ois {Schwarzentruber}}
  {5}
  {4th International Conference on Tools for Teaching Logic}
  {1}
  {1}
  {223}
\EventShortName{TTL2015}

\usepackage{graphicx}

\newtheorem{proposition}{Proposition}

\newcommand{\EUL}{{\sf EUL}}

\newcommand{\D}{\mathcal{D}}

\newcommand{\E}{\mathcal{E}}

\newcommand{\s}{\mathcal{S}} 

\newcommand{\nat}[1]{#1 ^{\circ}}

\newcommand{\imp}{\rightarrow}

\newcommand{\GDS}{{\sf GDS}}

\newcommand{\rel}{{\sf rel}} 

\newcommand{\namedpoint}[1]{\bullet \hspace{-1ex}\raisebox{1.2ex}{$#1 $}}

\newcommand{\tie}[2]{#1\bowtie #2}

\newcommand{\outof}[2]{#1\vdash\hspace*{-5pt}\dashv#2}

\newcommand{\inside}[2]{#1\sqsubset#2}

\newcommand{\allare}[2]{{\sf All $#1$ are $#2$}}
\newcommand{\noare}[2]{{\sf No $#1$ are $#2$}}
\newcommand{\someare}[2]{{\sf Some $#1$ are $#2$}}
\newcommand{\somenot}[2]{{\sf Some $#1$ are not $#2$}}

\begin{document}

\maketitle

\begin{abstract}
Understanding the notion of a model is not always easy in logic courses. 
Hence, tools such as Euler diagrams are frequently applied 
as informal illustrations of set-theoretical models. 
We formally investigate Euler diagrams as an introduction 
to set-theoretical models. 
We show that the model-theoretic notions of validity and invalidity 
are characterized by Euler diagrams, 
and, in particular, that model construction can be 
described as a manipulation of Euler diagrams. 
\end{abstract}

\section{Introduction}

Logic is traditionally studied from the different viewpoints of 
syntax and semantics. 
From the syntactic viewpoint, 
formal proofs are investigated using proof systems 
such as natural deduction and sequent calculus. 
From the semantic viewpoint, set-theoretical models of sentences 
are usually investigated. 
In contrast to a proof, which shows the validity of a given inference, 
we usually disprove an inference by constructing a counter-model. 
A counter-model is one in which 
all premises of a given inference are true, but its conclusion is false. 
The notions of proofs and models 
are traditionally defined in the fundamentally different frameworks 
of syntax and semantics, respectively, 
and the completeness theorem, 
one of the most basic theorems of logic, 
provides a bridge between them. 
In university courses, logic is usually taught along such lines. 

As the notion of a proof appears naturally in mathematics courses, 
students are, to some extent, familiar with it. 
However, the notion of a model can be difficult for beginners. 
A set-theoretical model consists of 
a domain of entities and sets, 
and an interpretation function that assigns truth values to sentences. 
Models are usually defined symbolically in the same way as syntax, and 
beginners often find it difficult 
to distinguish between syntax and semantics, 
as well as to understand the notion of the interpretation of symbols. 
Thus, some diagrams are used to introduce set-theoretical models. 
One famous example is 
Tarski's World of Barwise and Etchemendy \cite{Tarski'sworld}, 
which is designed to introduce model theoretic notions diagrammatically. 
Another traditional example concerns Euler diagrams, 
which were originally introduced in the 18th century to illustrate syllogisms. 
A basic Euler diagram consists of circles and points, 
and syllogistic sentences are represented 
using inclusion and exclusion relations between the circles and points. 
A circle in an Euler diagram 
can be considered to represent a set, and 
a point can be considered as an entity of a given domain. 
Inclusion and exclusion relations between circles and points 
can then be considered to represent 
set-theoretical notions such as subset and disjointness relations, 
respectively.

Besides the traditional informal semantic view, 
Euler diagrams have recently been investigated syntactically 
as the counterparts of logical formulas, 
which constitute formal proofs. 
Euler diagrams are rigorously defined as syntactic objects, 
allowing set-theoretical semantics to be defined. 
Inference systems are also formalized, 
and they have been shown to be equivalent to some symbolic logical systems. 
Consequently, fundamental logical properties such as 
soundness and completeness have been investigated 
(e.g., \cite{Shin,HowseSpider,MOT}). 
In this way, Euler diagrams are, 
on the one hand, regarded as models of given sentences, 
and on the other hand, 
they are exploited as syntactic objects that constitute formal proofs. 

In this paper, we present a formal investigation of Euler diagrams 
as set-theoretical models. 
We characterize the notions of validity and invalidity 
using our Euler diagrams. 
(In \cite{Barwise}, Barwise and Etchemendy also suggested 
a possible application of Euler diagrams to construct counter-examples.) 
The advantages of our view of Euler diagrams as models are as follows: 
(1) diagrams provide concrete images of set-theoretical models 
and counter-models; 
(2) diagrams provide natural images of the interpretation of sentences; 
(3) model and counter-model construction can be captured 
as concrete manipulations of diagrams. 
However, there are the following limitations: 
(1) the expressive power of our diagrams is restricted 
to conjunctions of extended syllogistic sentences; 
(2) not all sets are represented by the circles or simple closed curves 
of our diagrams; 
(3) diagrams are sometimes misleading 
if their meaning and manipulations are not precisely defined. 
Regardless, Euler diagrams have recently been defined rigorously 
from logical viewpoints, 
and we believe that this restricted fragment is sufficient 
for an introduction to model theory.

In Section \ref{sectionsyntactic}, 
we first review the Euler diagrammatic system of \cite{MOT}, 
in which Euler diagrams are regarded as syntactic objects 
corresponding to first-order formulas. 
Then, in Section \ref{sectionvalidity}, 
we investigate a view of Euler diagrams as 
counterparts of set-theoretical models. 
Finally, we investigate how Euler diagrams 
provide counter-models of given inferences 
in Section \ref{sectioncounter-models}.

\section{Euler diagrams as syntactic objects} \label{sectionsyntactic}

Our sentences have the following extended syllogistic form: 

\smallskip 
{\sf $a$ is $B$}, \ \ {\sf $a$ is not $B$}, \ \ 
{\sf There is something $B$}, \ \ {\sf There is something not $B$}, \ \ \\
\hspace*{1.5em} 
\allare{A}{B}, \ \noare{A}{B}, \ \someare{A}{B}, \ \somenot{A}{B}, 
\smallskip 

\noindent 
where $a$ is a constant and $A,B$ are predicates. 
These basic sentences are denoted by $S, S_1 ,S_2 ,\dots $, 
and we also consider their conjunctions. 

From a semantic viewpoint, 
every constant $a$ is interpreted as an element $I(a)$, 
and a predicate $B$ is interpreted as a {\it nonempty} set $I(B)$ 
in a set-theoretical domain. 
Then, our syllogistic sentences are interpreted as usual. 
For example, {\sf $a$ is $B$} is true iff $I(a)\in I(B)$; 
\allare{A}{B} is true iff $I(A)\subseteq I(B)$, and so on 
(cf. \cite{MOT}). 
Note that we impose the so-called {\it existential import}, 
i.e., the interpretation of $B$ is nonempty. 
See Remark 
after Example \ref{example5}.

\bigskip 

We first review the Euler diagrammatic system of \cite{MOT}, 
in which Euler diagrams are regarded as syntactic objects 
corresponding to formulas.

\begin{definition} \rm 
Our Euler diagram, called {\bf \EUL-diagram}, 
is defined as a plane with {\bf named circles} (simple closed curves, 
denoted by $A,B,C,\dots$) and {\bf named points} ($p,p_1 ,p_2 ,\dots$). 
Named points are divided into 
{\bf constant points} ($a,b,c,\dots $), which correspond to constants 
of the first-order language, 
and {\bf existential points} ($x,y,z,\dots $), 
which correspond to bound variables associated with 
the existential quantifier. 
Diagrams are denoted by $\D, \mathcal{E}, \D_1, \D_2, \dots $. \\ 
\end{definition}

Each diagram is specified by the following inclusion, exclusion, and 
crossing relations maintained between circles and points. 

\begin{definition} \rm 
{\bf \EUL-relations} consist of the following 
reflexive asymmetric binary relation $\inside{}{}$, and 
the irreflexive symmetric binary relations $\outof{}{}$ and $\tie{}{}$: 

\smallskip 

\begin{tabular}{cl} 
$\inside{A}{B}$  &  \normalsize ``the interior 
of $A$ is {\it inside of} the interior of $B$,''  \\

$\outof{A}{B}$ & \normalsize 
``the interior of $A$ is {\it outside of} the interior of $B$,'' \\
$\tie{A}{B}$  &  \normalsize 
``there is at least one {\it crossing} point between $A$ and $B$,''  \\

$\inside{p}{A}$  &  \normalsize 
``$p$ is {\it inside of} the interior of $A$,'' \\ 

$\outof{p}{A}$  &  \normalsize 
``$p$ is {\it outside of} the interior of $A$,'' \\ 

$\outof{p}{q}$  &  \normalsize 
``$p$ is {\it outside of} $q$ 
(i.e., $p$ is not located at point $q$).'' 
\end{tabular} 

\smallskip \noindent 
\EUL-relations are denoted by $R, R_1 ,R_2 ,\dots $. 
\end{definition}

The set of \EUL-relations that hold on a diagram $\D$ is uniquely determined, 
and we denote this set by $\rel(\D)$. 
For example, $\rel(\s_1 +\s_2 )$ in Example \ref{examplesyllog} 
is $\{ \inside{x}{A}, \inside{x}{B}, \inside{x}{C}, \tie{A}{B},$ 
$\tie{A}{C}, \inside{B}{C} \}$. 
In the description of $\rel(\D)$, 
we omit the reflexive relation $\inside{s}{s}$ for each object $s$. 
Furthermore, we often omit relations of the form $\outof{p}{q}$ 
for points $p$ and $q$, 
which always hold by definition. 
In the following, 
we consider the equivalence class of diagrams in terms of the \EUL-relations.

To clarify the intended meaning of our diagrams, 
we describe the translation of diagrams into the usual first-order formulas. 
See \cite{EulerNormal,counterdiag} 
for a detailed description of our translation. 
Each named circle is translated into a unary predicate, 
and each constant (resp. existential) point is translated into 
a constant symbol (resp. variable). 
Then, each \EUL-relation $R$ is translated into 
a formula $\nat{R}$ as follows: 
\begin{itemize} 
\item[] $\nat{(\inside{p}{A})} := A(p)$; \hspace*{8em} 
$\nat{(\outof{p}{A})} := \neg A(p)$; 
\item[] $\nat{(\inside{A}{B})} := \forall x(A(x)\imp B(x))$; \hspace{2em} 
$\nat{(\outof{A}{B})} := \forall x(A(x)\imp \neg B(x))$; \hspace{2em} 
\item[] $\nat{(\tie{A}{B})} := \forall x((A(x)\imp A(x))\wedge (B(x)\imp B(x)))$. 
\end{itemize} 

\noindent 
Let $\D$ be an \EUL-diagram with 
the following set of relations: 
$\rel(\D) = \{ R_1 ,\dots ,R_i , 
x_1 \Box A_1 ,$
$\dots ,x_1 \Box A_k , \dots ,x_l \Box A_1 ,\dots ,x_l \Box A_k \},$
where $\Box$ is $\inside{}{}$ or $\outof{}{}$, and 
no existential point appears in $R_1 ,\dots ,R_i$. 
Then, the diagram $\D$ is translated into the following conjunctive formula: 
$$
\nat{\D} := 
\nat{R_1} \wedge \cdots \wedge \nat{R_i} \wedge 
\exists x_1 (\overline{A_1 (x_1 )}\wedge \cdots \wedge \overline{A_k (x_1 )}) 
\wedge \cdots \wedge 
\exists x_l (\overline{A_1 (x_l )}\wedge \cdots \wedge \overline{A_k (x_l )}), 
$$
where $\overline{A(x)}$ is $A(x)$ or $\neg A(x)$ depending on $\Box$.

Note that our diagram is abstractly a ``conjunction of relations.''
Note also that we interpret the $\tie{}{}$-relation so that 
it does not convey any specific (inclusion or exclusion) information 
about the relationship between circles. 
(This interpretation is based on the convention of Venn diagrams. 
See \cite{MOT} for further details.) 
Thus, $\tie{A}{B}$ is translated into a tautology as above.

From a semantic viewpoint, 
by interpreting circles (resp. points) as 
{\it nonempty} subsets (resp. elements) 
of a set-theoretical domain, 
each ``syntactic object'' \EUL-diagram is interpreted 
in terms of the relations that hold on it. 
See \cite{MOT} for details.

\begin{remark}\rm 
Our system is essentially the same as the {\it region connection calculus RCC} 
\cite{RCC1992}. 
RCC is a topological approach 
to qualitative spatial representation and reasoning, 
and is applied, for example, to Geographic Information System (GIS). 
RCC investigates eight basic relations, 
including our inclusion, exclusion, and crossing (partially overlapping), 
between spatial regions (circles in our framework). 
Although RCC investigates general $n$-dimensional spaces of spatial regions, 
we concentrate on 2-dimensional diagrams. 
Thus, without any named points, our system can be considered 
as a subsystem of RCC. 
See, for example, \cite{CohnHazarika2001,CohnRenz2008} 
for surveys of RCC. 
\end{remark}

We next review the Euler diagrammatic inference system of 
\cite{MOT}, 
called the Generalized Diagrammatic Syllogistic inference system \GDS. 
\GDS\ consists of two kinds of inference rules: 
{\sf Deletion} and {\sf Unification}. 
{\sf Deletion} allows us to delete a circle or point 
from a given diagram. 
{\sf Unification} allows us to unify two diagrams into one diagram 
{\it in which the semantic information is equivalent to 
the conjunction of the original two diagrams.} 

\begin{example} \label{examplesyllog} \rm 
\someare{A}{B}, \allare{B}{C} $\models$ \someare{A}{C} 
\end{example} 

\noindent 
\raisebox{2em}{
\begin{minipage}[b]{25em} 
The validity of the given inference is shown 
by the {\sf Unification} of the two diagrams $\s_1 $ and $\s_2$, 
which represent the given premises. 
In this unification, 
circle $B$ in $\s_1 $ and $\s_2 $ is identified, 
and $C$ is added to $\s_1 $ so that $B$ is inside $C$ 
and $C$ overlaps with $A$ {\it without any implication of 
a specific relationship between $A$ and $C$}. 
$\s_3$, which represents the given conclusion, 
is then obtained by deleting $B$. 
\end{minipage} 
}
\hspace*{1em} 
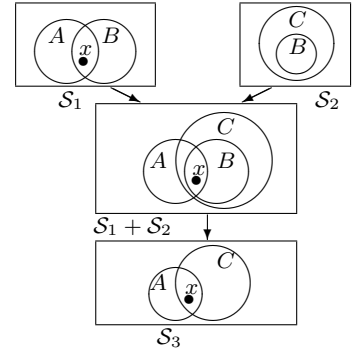
\begin{minipage}[t]{15em} 
\begin{center} \unitlength=1.2pt \footnotesize \begin{picture}(100,90) 
\put(-10,80){\unitlength=1pt \fbox{
 \begin{picture}(43,25)
 \put(13,10){\circle{23}} \put(6,13){$A$}
 \put(17,4){$\namedpoint{x}$}
 \put(27,10){\circle{23}} \put(26,13){$B$} 
 \put(10,-11){$\s_1$} 
 \end{picture}}}
\put(60,80){\unitlength=1pt \fbox{\begin{picture}(35,25)
 \put(18,9){\circle{15}} \put(15,9){$B$}
 \put(18,13){\circle{27}} \put(15,19){$C$}
 \put(25,-11){$\s_2$} 
 \end{picture}}} 
\put(20,77){\vector(2,-1){9}} 
\put(70,77){\vector(-2,-1){9}} 
\put(15,40){\unitlength=1.4pt \fbox{ 
 \begin{picture}(45,25)
 \put(28,9){\circle{17}} \put(28,10){$B$}
 \put(30,12){\circle{26}} \put(28,19){$C$}
 \put(17,9){\circle{18}} \put(10,10){$A$} 
 \put(21,5){$\namedpoint{x}$} 
 \put(-5,-7.5){$\s_1 +\s_2$} 
 \end{picture} }} 
\put(50,37){\vector(0,-1){8}} 
\put(15,5){\unitlength=1.4pt \fbox{ 
 \begin{picture}(45,18)
 \put(27,9){\circle{20}} \put(28,13){$C$}
 \put(17,6){\circle{15}} \put(10,7){$A$} 
 \put(19,3){$\namedpoint{x}$} 
 \put(12,-7.5){$\s_3$} 
 \end{picture} }} 
\end{picture} 
\vspace*{1em} 
\end{center} 
\end{minipage}

Two types of constraint are imposed on {\sf Unification}. 
One is the {\it constraint for determinacy}, 
which blocks disjunctive ambiguity with respect to 
the location of points, and 
the other is the {\it constraint for consistency}, 
which blocks representing inconsistent information 
in a single diagram. 
{\sf Unification} can only be applied when 
these constraints are satisfied.

The notion of {\bf diagrammatic proof}, or {\bf d-proof} for short, 
is defined inductively as a tree structure consisting of 
{\sf Unification} and {\sf Deletion} steps. 
The provability relation $\vdash$ 
is defined as usual in terms of the existence of a d-proof. 
See \cite{MOT,Soken2013} for details.

\GDS\ is shown to be sound and complete with respect to 
our set-theoretical semantics. 
To avoid introducing the diagrammatic counterpart of 
the so-called absurdity rule in our system, 
we impose a consistency condition on the set of premise diagrams 
$\D_1 ,\dots ,\D_n$ 
in our formulation of completeness, 
i.e., the set of premise diagrams has a model. 
See \cite{MOT,Soken2013} for further discussion and a detailed proof.

\begin{proposition} \label{completeness}
Let $\D_1 ,\dots ,\D_n $ be consistent.  
$\E$ is a semantic consequence of $\D_1 ,\dots ,\D_n$ 
($\D_1 ,\dots ,\D_n \models \E$) if and only if 
$\E$ is provable from $\D_1 ,\dots ,\D_n$ 
($\D_1 ,\dots ,\D_n \vdash \E$) in \GDS. 
\end{proposition}

Note that the above completeness is obtained 
by regarding diagrams as ``syntactic objects'', 
corresponding to first-order formulas, 
and by defining appropriate set-theoretical semantics. 
We next investigate a view of 
Euler diagrams as counterparts of set-theoretical models.

\section{Euler diagrams as models} \label{sectionvalidity}

By regarding our diagrams as models, 
we define the truth condition for our sentences. 

\begin{definition} \rm 
Constants are interpreted as named points, 
predicates are circles. 
Then, for a diagram $\D$: 
\begin{itemize} 
\item {\sf $a$ is $B$} holds on $\D$ iff 
named point $a$ is located inside circle $B$; 
\item {\sf $a$ is not $B$} holds on $\D$ iff 
named point $a$ is located outside circle $B$; 
\item \allare{A}{B} holds on $\D$ iff $\inside{A}{B}$ holds on $\D$; 
\item \noare{A}{B} holds on $\D$ iff $\outof{A}{B}$ holds on $\D$; 
\item \someare{A}{B} holds on $\D$ iff 
there exists an existential point $x$ in the intersection of $A$ and $B$; 
\item \somenot{A}{B} holds on $\D$ iff 
there exists an existential point $x$ inside $A$ and outside $B$. 
\end{itemize} 
\noindent 
We write $\D \Vdash S$ when sentence $S$ holds on diagram $\D$. 
\end{definition} 

\begin{example} \label{examplesomeare} \rm 
\someare{A}{B} holds on each of the following diagrams, for example. 
\end{example} 
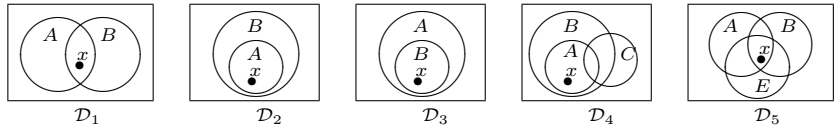
\begin{center} \unitlength=1.2pt \scriptsize 
\fbox{\begin{picture}(40,25)
\put(13,12){\circle{23}} \put(8,16){$A$} 
\put(27,12){\circle{23}} \put(26,16){$B$} 
\put(18,7){$\namedpoint{x}$} 
\put(18,-9){$\D_1$} 
\end{picture}} 
\hspace*{1em} 
\fbox{\begin{picture}(35,25)
\put(18,8){\circle{16}} \put(15,10){$A$}
\put(18,12){\circle{26}} \put(15,19){$B$} 
\put(15,2){$\namedpoint{x}$} 
\put(18,-9){$\D_2$} 
\end{picture}} 
\hspace*{1em} 
\fbox{\begin{picture}(35,25)
\put(18,8){\circle{16}} \put(15,10){$B$}
\put(18,12){\circle{26}} \put(15,19){$A$} 
\put(15,2){$\namedpoint{x}$} 
\put(18,-9){$\D_3$} 
\end{picture}} 
\hspace*{1em} 
\fbox{\begin{picture}(35,25)
\put(13,8){\circle{16}} \put(10,11){$A$}
\put(13,12){\circle{26}} \put(10,19){$B$} 
\put(10,2){$\namedpoint{x}$} 
\put(25,10){\circle{18}} \put(28,10){$C$} 
\put(18,-9){$\D_4$} 
\end{picture}} 
\hspace*{1em} 
\fbox{\begin{picture}(40,25)
\put(14,15){\circle{20}} \put(8,19){$A$} 
\put(26,15){\circle{20}} \put(26,19){$B$} 
\put(18.5,9){$\namedpoint{x}$} 
\put(19,8){\circle{19}} \put(18,0){$E$} 
\put(18,-9){$\D_5$} 
\end{picture}} 
\vspace*{1em} 
\end{center}

We define the notion of {\bf Euler diagrammatic validity} 
in the same way as model-theoretic validity. 

\begin{definition} \label{Eulervalidity} \rm 
For $n\neq 0$, 
$S_1 ,\dots ,S_n \Vdash S$ if $S$ holds on any diagram $\D$ 
on which $S_1 ,\dots ,S_n $ hold 
(i.e., $\forall \D. \D \Vdash \bigwedge S_i \Rightarrow \D \Vdash S$). 
\end{definition}

In our {\sf Unification} of \GDS, 
as illustrated in Example \ref{examplesyllog}, 
when the relationship between circles cannot be determined 
as $\inside{}{}$ or $\outof{}{}$ from the given premises, 
we assign this to be a $\tie{}{}$-relation without any implication. 
However, if we extend the notion of ``unification'' 
to allow additional implications 
(and hence, such that this ``unification'' is invalid) 
as long as all relations on premises are preserved, 
then we can describe model construction 
in terms of the extended ``unification'' as follows.

\begin{example}\rm 
\someare{A}{B}, \allare{B}{C} $\Vdash$ \someare{A}{C} 

\noindent 
Various diagrams can represent the given premises, 
and there are various ways to ``unify'' these diagrams 
(using {\sf Deletion} if necessary). 
The following describes three possible diagrams, 
and the conclusion \someare{A}{C} holds in each of them. 
Cf. Example \ref{examplesyllog}, 
where the representations of sentences are fixed canonically 
in a one-to-one correspondence depending on each sentence, 
and the way of unification is also fixed uniquely. 
\end{example} 
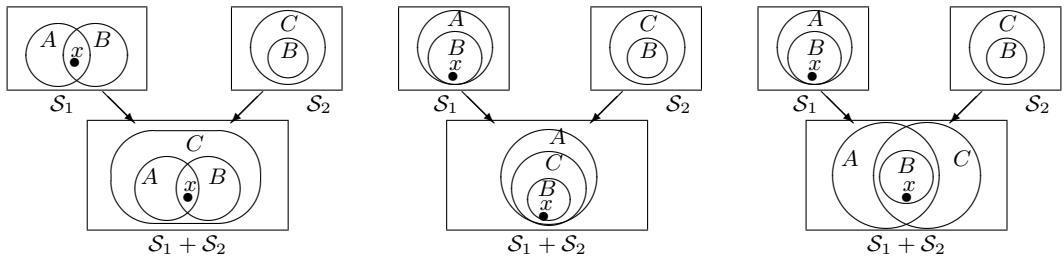
\begin{center} \unitlength=1.2pt \footnotesize 
\begin{picture}(100,70) 
\put(-10,44){\unitlength=1pt \fbox{
 \begin{picture}(43,25)
 \put(13,10){\circle{23}} \put(6,13){$A$}
 \put(17,5){$\namedpoint{x}$}
 \put(27,10){\circle{23}} \put(26,13){$B$} 
 \put(10,-11){$\s_1$} 
 \end{picture}}}
\put(60,44){\unitlength=1pt \fbox{\begin{picture}(35,25)
 \put(18,9){\circle{15}} \put(15,9){$B$}
 \put(18,13){\circle{27}} \put(15,19){$C$}
 \put(25,-11){$\s_2$} 
 \end{picture}}} 
\put(20,41){\vector(1,-1){10}} 
\put(70,41){\vector(-1,-1){10}} 
\put(15,0){\unitlength=1.4pt \fbox{ 
 \begin{picture}(45,25) 
 \put(28,9){\circle{17}} \put(28,10){$B$} 
 \put(22,12){\oval(40,25)} \put(22,19){$C$}
 \put(17,9){\circle{18}} \put(10,10){$A$} 
 \put(21,5){$\namedpoint{x}$} 
 \put(12,-8){$\s_1 +\s_2$} 
 \end{picture} }} 
\end{picture} 
\hspace*{1em} 
\begin{picture}(100,50) 
\put(0,44){\unitlength=1pt \fbox{\begin{picture}(35,25)
 \put(18,9){\circle{18}} \put(15,10){$B$}
 \put(18,13){\circle{29}} \put(15,21){$A$} 
 \put(15,0){$\namedpoint{x}$} 
 \put(10,-11){$\s_1$} 
 \end{picture}}}
\put(60,44){\unitlength=1pt \fbox{\begin{picture}(35,25)
 \put(18,9){\circle{15}} \put(15,9){$B$}
 \put(18,13){\circle{27}} \put(15,19){$C$}
 \put(25,-11){$\s_2$} 
 \end{picture}}} 
\put(20,41){\vector(1,-1){10}} 
\put(70,41){\vector(-1,-1){10}} 
\put(15,0){\unitlength=1.4pt \fbox{ 
 \begin{picture}(45,25)
 \put(23,9){\circle{20}} \put(22,14){$C$}
 \put(23,12){\circle{27}} \put(23,21){$A$} 
 \put(23,6){\circle{12}} \put(20,7){$B$} 
 \put(20,0){$\namedpoint{x}$} 
 \put(12,-8){$\s_1 +\s_2$} 
 \end{picture} }} 
\end{picture} 
\hspace*{1em} 
\begin{picture}(100,50) 
\put(0,44){\unitlength=1pt \fbox{\begin{picture}(35,25)
 \put(18,9){\circle{18}} \put(15,10){$B$}
 \put(18,13){\circle{29}} \put(15,21){$A$} 
 \put(15,0){$\namedpoint{x}$} 
 \put(10,-11){$\s_1$} 
 \end{picture}}}
\put(60,44){\unitlength=1pt \fbox{\begin{picture}(35,25)
 \put(18,9){\circle{15}} \put(15,9){$B$}
 \put(18,13){\circle{27}} \put(15,19){$C$}
 \put(25,-11){$\s_2$} 
 \end{picture}}} 
\put(20,41){\vector(1,-1){10}} 
\put(70,41){\vector(-1,-1){10}} 
\put(15,0){\unitlength=1.4pt \fbox{ 
 \begin{picture}(45,25) 
 \put(28,12.5){\circle{27.5}} \put(35,15){$C$} 
 \put(17,12.5){\circle{27.5}} \put(5,15){$A$} 
 \put(22.5,12){\circle{14}} \put(20,13){$B$} 
 \put(21,5){$\namedpoint{x}$} 
 \put(12,-8){$\s_1 +\s_2$} 
 \end{picture} }} 
\end{picture} 
\vspace*{1em} 
\end{center} 

It is shown that the canonical form of the diagram for every sentence 
(e.g., $\D_1$ in Example \ref{examplesomeare} for \someare{A}{B}) 
is sufficient to consider the usual model-theoretic validity. 
Then, through the completeness of \GDS, 
it is shown that Euler diagrammatic validity of Definition \ref{Eulervalidity} 
is equivalent to the usual model-theoretic validity.

\section{Euler diagrams for counter-models} \label{sectioncounter-models}

Models can be practically applied to disprove given inferences. 
Let us consider how to disprove a given inference 
using Euler diagrams.

\begin{example} \label{example1} \rm 
{\sf Some $A$ are $B$}, {\sf All $B$ are $C$} 
$\not \models$ {\sf All $A$ are $C$} 
\end{example} 
\noindent 
\begin{minipage}[b]{28em} 
These premises are the same as in Example \ref{examplesyllog}. 
To construct a diagram to falsify 
the given conclusion {\sf All $A$ are $C$}, 
we add a fresh existential point $y$ to the unified diagram $\s_1 +\s_2$ 
of Example \ref{examplesyllog} as follows. 
\end{minipage} 
\hspace*{1em} 
\raisebox{1em}{ 
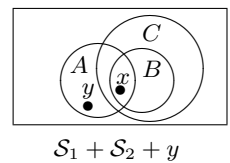
\begin{minipage}[t]{10em} \unitlength=1.5pt \small \fbox{ 
 \begin{picture}(45,25)
 \put(28,9){\circle{17}} \put(28,10){$B$}
 \put(30,12){\circle{26}} \put(28,19){$C$}
 \put(17,9){\circle{18}} \put(10,11){$A$} 
 \put(21,5){$\namedpoint{x}$} 
 \put(13,5.5){$y$} 
 \put(13,1){$\bullet$} 
 \put(6,-10){$\s_1 +\s_2 +y$} 
 \end{picture} } 
\end{minipage} 
}

\noindent 
The existential point $y$ denotes that 
``there exists something that is $A$ but not $C$,'' 
i.e., the negation of {\sf All $A$ are $C$}. 
Note that even after the addition of $y$, 
all of the $\inside{}{},\outof{}{}$ relations that hold on the premises 
($\inside{x}{A}$ and $\inside{x}{B}$ on 
$\s_1$, and $\inside{B}{C}$ on $\s_2$) 
are maintained. 
Hence, the above $\s_1 +\s_2 +y$ neatly illustrates a counter-model 
of the given inference 
in which all premises are true but the conclusion is false.

Let us define our counter-diagrams.

\begin{definition} \rm 
A diagram $\E$ is a {\bf counter-diagram} of $\D$, and vice versa, 
when one of the following holds between $\D$ and $\E$: 
\begin{itemize} \setlength \itemsep{0mm} 
\item $\inside{a}{B}$ holds on $\D$, and $\outof{a}{B}$ holds on $\E$; 
 \item $\inside{A}{B}$ holds on $\D$, and, for some $x$, 
$\inside{x}{A} \mbox{ and } \outof{x}{B}$ hold on $\E$; 
 \item $\outof{A}{B}$ holds on $\D$, and, for some $x$, 
$\inside{x}{A} \mbox{ and } \inside{x}{B}$ hold on $\E$. 
\end{itemize} 
\end{definition}

Note that our diagram is abstractly a conjunction of relations, 
and the above definition lists 
the so-called ``counter-relation'' for every relation. 
As illustrated in Example \ref{examplecounter-diagrams}, 
there may be a number of counter-diagrams for a given diagram $\D$, 
because a counter-diagram of $\D$ may contain circles and points 
that are irrelevant to $\D$. 
Thus, in general, the notions of a counter-diagram of $\D$ and 
the negation of $\D$, which should be uniquely determined, 
are different.

\begin{example} \label{examplecounter-diagrams} \rm 
Let $\D$ be the following diagram on the left. 
Then, all of $\D_6 ,\D_7 ,\D_8$, and $\D_9$ 
are counter-diagrams of $\D$. 
Thus, 
a diagram and its counter-diagram do not necessarily 
share the same circles and points, and 
it is sufficient that a relation that holds on a counter-diagram 
falsifies a relation of the given diagram. 
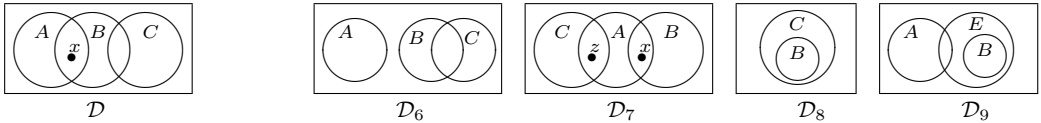
\begin{center} \unitlength=1.1pt \scriptsize 
\fbox{\begin{picture}(58,25)
\put(13,12){\circle{24}} \put(7,16){$A$}
\put(18,8){$\namedpoint{x}$}
\put(27,12){\circle{24}} \put(26,16){$B$}
\put(45,12){\circle{24}} \put(44,16){$C$} 
\put(25,-11){\footnotesize $\D$} 
\end{picture}} 
\hspace*{5em} 
\fbox{\begin{picture}(58,25)
\put(11,12){\circle{22}} \put(5,16){$A$}
\put(37,12){\circle{20}} \put(29,14){$B$}
\put(48,12){\circle{20}} \put(48,14){$C$} 
\put(25,-11){\footnotesize $\D_6$} 
\end{picture}} 
\hspace*{1ex} 
\fbox{\begin{picture}(58,25)
\put(13,12){\circle{24}} \put(7,16){$C$}
\put(18,8){$\namedpoint{z}$}
\put(28,12){\circle{25}} \put(26,16){$A$}
\put(35,8){$\namedpoint{x}$} 
\put(45,12){\circle{24}} \put(44,16){$B$} 
\put(25,-11){\footnotesize $\D_7$} 
\end{picture}} 
\hspace*{1ex} 
\fbox{\begin{picture}(35,25)
\put(18,9){\circle{15}} \put(15,9){$B$}
\put(18,13){\circle{27}} \put(15,19){$C$}
\put(18,-11){\footnotesize $\D_8$} 
\end{picture}} 
\hspace*{1ex} 
\fbox{\begin{picture}(48,25)
\put(11,12){\circle{22}} \put(5,16){$A$}
\put(33,10){\circle{15}} \put(30,10){$B$}
\put(30,12){\circle{27}} \put(27,19){$E$} 
\put(25,-11){\footnotesize $\D_9$} 
\end{picture}} 
\vspace*{1em} 
\end{center} 
\end{example}

\bigskip 

We can list all of the 
operations to construct counter-diagrams, 
such as the addition of a new point illustrated in Example \ref{example1}. 
However, let us consider another way to construct counter-diagrams. 
Theoretically speaking, 
$S_1 ,\dots ,S_n \not \Vdash T$ means that 
there exists a $\D$ such that 
$\D \Vdash S_1 \wedge \cdots \wedge S_n \wedge \neg T$. 
Based on this fact, 
we construct a diagram that falsifies the given conclusion, 
and, by unifying the diagram with all premises, 
we construct a counter-diagram for the given inference. 
We formalize our counter-diagram construction according to this idea. 
Let us reconsider the inference in Example \ref{example1}. 

\begin{example} \label{example1-2} \rm 
{\sf Some $A$ are $B$}, {\sf All $B$ are $C$} 
$\not \models$ {\sf All $A$ are $C$} 

\noindent 
Let $\E'$ be a diagram such that 
$\rel(\E')=\{ \inside{y}{A}, \outof{y}{C}, \tie{A}{C} \}$, 
which represents {\it Some $A$ are not $C$}, and 
which falsifies the given conclusion {\sf All $A$ are $C$}. 
We unify this $\E'$ with $\s_1$ and $\s_2$, 
as illustrated in Fig.\,\ref{figureexample1-2}. 
Thus, the resulting diagram $\s_1 +\s_2 +\E'$ is the same as 
$\s_1 +\s_2 +y$ in Example \ref{example1}, 
and disproves the given inference. 
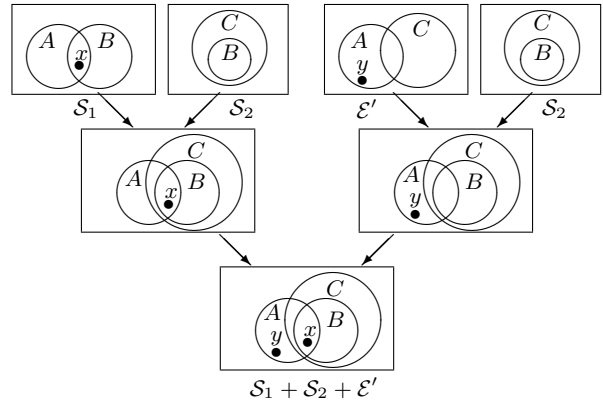
\begin{figure}[!h] 
\begin{center} \unitlength=1.3pt \footnotesize 
\begin{picture}(100,110) 
\put(-45,80){\unitlength=1.1pt 
 \fbox{\begin{picture}(43,25)
 \put(13,10){\circle{23}} \put(6,13){$A$}
 \put(18,5){$\namedpoint{x}$}
 \put(27,10){\circle{23}} \put(26,13){$B$} 
 \put(18,-10){$\s_1$} 
 \end{picture}}}
\put(0,80){\unitlength=1.1pt 
 \fbox{\begin{picture}(35,25)
 \put(18,9){\circle{15}} \put(15,9){$B$}
 \put(18,13){\circle{27}} \put(15,19){$C$}
 \put(18,-10){$\s_2$} 
 \end{picture}}} 
\put(-20,78){\vector(1,-1){10}} 
\put(15,78){\vector(-1,-1){10}} 
\put(45,80){\unitlength=1.1pt 
 \fbox{\begin{picture}(43,25)
 \put(13,10){\circle{23}} \put(6,13){$A$}
 \put(8,5.5){$y$} 
 \put(8,0){$\bullet$} 
 \put(29,12){\circle{27}} \put(27,17){$C$} 
 \put(8,-11){$\E'$} 
 \end{picture}}}
\put(90,80){\unitlength=1.1pt 
 \fbox{\begin{picture}(35,25)
 \put(18,9){\circle{15}} \put(15,9){$B$}
 \put(18,13){\circle{27}} \put(15,19){$C$}
 \put(18,-10){$\s_2$} 
 \end{picture}}} 
\put(65,78){\vector(1,-1){10}} 
\put(105,78){\vector(-1,-1){10}} 
\put(-25,40){\fbox{\begin{picture}(45,25)
 \put(28,9){\circle{17}} \put(28,10){$B$}
 \put(30,12){\circle{27}} \put(28,19){$C$}
 \put(17,9){\circle{18}} \put(10,11){$A$} 
 \put(21,4){$\namedpoint{x}$} 
 \end{picture}}} 
\put(55,40){\fbox{\begin{picture}(45,25)
 \put(28,9){\circle{17}} \put(28,10){$B$}
 \put(30,12){\circle{27}} \put(28,19){$C$}
 \put(17,9){\circle{19}} \put(10,12){$A$} 
 \put(12,6){$y$} 
 \put(12,1){$\bullet$} 
 \end{picture}}} 
\put(15,37){\vector(1,-1){9}} 
\put(65,37){\vector(-1,-1){9}} 
\put(15,0){\fbox{\begin{picture}(45,25)
 \put(28,9){\circle{17}} \put(28,10){$B$}
 \put(30,12){\circle{27}} \put(28,19){$C$}
 \put(17,9){\circle{19}} \put(10,12){$A$} 
 \put(21,4){$\namedpoint{x}$} 
 \put(12,6){$y$} 
 \put(12,1){$\bullet$} 
 \put(6,-10){$\s_1 +\s_2 +\E'$} 
 \end{picture}}} 
\end{picture} 
\vspace*{1em} 
\end{center} 
\vspace*{-1em} 
\caption{Construction of a counter-diagram (Example \ref{example1-2})} \label{figureexample1-2} 
\end{figure} 
\vspace*{1ex} 
\end{example}

Let us examine another example.

\begin{example} \label{example5} \rm 
{\sf No $A$ are $B$}, {\sf There is something $C$} $\not \models$ 
{\sf No $A$ are $C$} 

\noindent 
Let $\E'$ be a diagram such that 
$\rel(\E')=\{ \inside{y}{A}, \inside{y}{C}, \tie{A}{C} \}$, 
which represents {\sf Some $A$ are $C$}, and 
which falsifies the given conclusion {\sf No $A$ are $C$}. 
We then unify this $\E'$ with the given premise $\s_3$, 
representing {\sf No $A$ are $B$}, 
as illustrated in Fig.\,\ref{figureexample5}. 
When we further attempt to unify diagram $\s_4$, 
in which $\{ \inside{x}{C} \}$ holds, 
representing {\sf There is something $C$}, 
we find three possibilities for the position of $x$, 
that is, $x$ is indeterminate with respect to circles $A$ and $B$, 
and we cannot unify this from our valid {\sf Unification} rules. 
However, whichever of the three possibilities 
gives the position of $x$, 
we obtain a counter-diagram to disprove the given inference. 
For example, $\E''$ in Fig.\,\ref{figureexample5} 
is a required counter-diagram. 
\end{example}

\begin{figure}[h] 
\begin{center} \unitlength=1.2pt \small 
\hspace*{-15em} 
\begin{picture}(100,81) 
\put(-10,48){\fbox{\begin{picture}(43,21)
 \put(13,10){\circle{22}} \put(6,13){$A$}
 \put(18,10){$y$} 
 \put(18,4){$\bullet$} 
 \put(27,10){\circle{22}} \put(26,13){$C$} 
 \put(18,-11){$\E'$} 
 \end{picture}}}
\put(60,48){\fbox{\begin{picture}(40,20)
  \put(10,10){\circle{18}} \put(5,12){$A$}
  \put(30,10){\circle{18}} \put(27,12){$B$} 
  \put(20,-10){$\s_3$} \end{picture}}} 
\put(150,48){\fbox{\begin{picture}(30,20)
 \put(14,10){\circle{20}} \put(11,13){$C$}
 \put(12,2){$\namedpoint{x}$} 
 \put(12,-10){$\s_4$} \end{picture}}} 
\put(20,45){\vector(1,-1){10}} 
\put(70,45){\vector(-1,-1){10}} 
\put(20,10){\fbox{\begin{picture}(53,21)
 \put(13,10){\circle{22}} \put(6,13){$A$}
 \put(18,10){$y$} 
 \put(18,4){$\bullet$} 
 \put(27,10){\circle{22}} \put(26,13){$C$} 
 \put(43,10){\circle{21}} \put(43,13){$B$} 
 \put(18,-11){$\E' +\s_3$} 
 \end{picture}}} 
\put(200,10){\fbox{\begin{picture}(53,21)
 \put(13,10){\circle{22}} \put(6,13){$A$}
 \put(18,10){$y$} 
 \put(18,4){$\bullet$} 
 \put(27,10){\circle{22}} \put(26,13){$C$} 
 \put(25,1){$\namedpoint{x}$} 
 \put(43,10){\circle{21}} \put(43,13){$B$} 
 \put(18,-11){$\E''$} 
 \end{picture}}} 
\end{picture} 
\end{center} 
\vspace*{-2em} 
\caption{(Example \ref{example5})} \label{figureexample5} 
\end{figure} 

\noindent 
Because we cannot obtain the above $\E''$ 
from $\E'+\s_3$ and $\s_4$ using our valid {\sf Unification} rules of \GDS, 
we construct counter-diagrams 
by introducing another ``invalid'' rule: {\sf invalid Point Insertion (iPI)}. 
This arbitrarily fixes the position of a point from several possibilities. 
Then, we obtain $\E''$ from $\E' +\s_3$ and $\s_4$.

In this way, a given inference is shown to be invalid by: 
(1) constructing a diagram that falsifies the given conclusion; and 
(2) unifying the diagram of (1) with all premises. 
In the resulting diagram, 
the relation falsifies the given conclusion and 
all of the $\inside{}{}$ and $\outof{}{}$ relations 
of the given premises hold. 
Hence, it is a counter-diagram for the given inference.

\begin{remark} \label{remarkexistentialimport} \rm 
In our system, we postulate the so-called {\it existential import} 
in the literature of syllogisms. 
With this postulate, for example, 
we have \allare{A}{B}, \noare{B}{C} $\models$ \somenot{A}{C}, 
even though this is not valid from the usual logical viewpoint 
without the existential import. 
Without this postulate, 
two diagrams, say $\D$ in which $\inside{A}{B}$ holds 
and $\E$ in which $\outof{A}{B}$ holds, are consistent 
when $A$ denotes the empty set. 
However, it is difficult to express $\D$ and $\E$ in a single diagram, 
as our system lacks a device to express the emptiness of circles. 
Even if we introduce another device such as ``shading'', 
which expresses the emptiness of the corresponding region (cf. \cite{Shin}), 
there remains a question as to whether we draw the shaded circle $A$ 
inside or outside $B$ (whichever is legal in an abstract sense). 
Thus, it seems that the assumption of the existential import 
is natural for our Euler diagrammatic system. 
\end{remark}

Let us define our counter-diagrammatic proofs. 

\begin{definition} \rm 
A {\bf counter-diagrammatic proof}, or {\bf counter-d-proof} for short, 
of $\E$ under $\D_1 ,\dots ,\D_n$ is a tree structure: 
\begin{itemize} \setlength \itemsep{0mm} 
\item consisting of valid {\sf Unification} rules 
and invalid {\sf iPI}; 
\item whose premises are $\D_1 ,\dots ,\D_n , \E'$, 
where $\E'$ is a counter-diagram of $\E$ such that 
no existential point of $\E'$ appears on $\D_1 ,\dots ,\D_n$; 
\item in whose conclusion, 
all $\inside{}{},\outof{}{}$-relations of $\D_1 ,\dots ,\D_n , \E'$ hold. 
\end{itemize} 

\noindent 
We write $\D_1 ,\dots ,\D_n \dashrightarrow \E$ 
when there exists a counter-d-proof of $\E$ under $\D_1 ,\dots ,\D_n$. 
\end{definition}

Note that we may not use the invalid {\sf iPI} 
as illustrated in Example \ref{example1-2}. 
Additionally, we can freely choose a counter-diagram $\E'$ 
to add to the premises of a counter-d-proof. 
Hence, it may be the case that $\E'$ is itself a required conclusion 
of a counter-d-proof. 
In our proof of Proposition \ref{theoremcounterproof} below, 
we present a canonical method to choose $\E'$ and 
construct counter-d-proofs. 
See \cite{counterdiag} for further details.

Although space limitations prevent us from going into detail, 
there exists an invalid inference 
for which we cannot construct a counter-d-proof in our framework 
of inclusion, exclusion, and crossing relations. 
Thus, we need to restrict the position of existential points 
appearing in a given conclusion. 
We call a diagram ``relational'' if the position of 
every existential point in the diagram is determined by relations 
with at most two circles. 
See \cite{counterdiag} for a detailed discussion. 
Thus, we have shown that our counter-d-proofs sufficiently 
characterize the notion of invalidity.

\begin{proposition}\label{theoremcounterproof} 
Let $\D_1 ,\dots ,\D_n $ be consistent, and $\E$ be relational. 
$\E$ is not a valid conclusion of $\D_1 ,\dots ,\D_n$ 
(i.e., $\D_1 ,\dots ,\D_n \not \models \E$) if and only if 
there exists a counter-d-proof of $\E$ under $\D_1 ,\dots ,\D_n$ 
(i.e., $\D_1 ,\dots ,\D_n \dashrightarrow \E$). 
\end{proposition}

There is no need to worry about relational diagrams 
when we consider the usual syllogisms or transitive inference, 
where a conclusion diagram consists of only two circles, 
and it is already relational. 
We require the restriction 
when we consider a more general diagram as a conclusion, 
in which only a particular existential point makes 
the given inference invalid. See \cite{counterdiag}.






\newpage
\thispagestyle{empty}
{\ }

\end{document}